\RequirePackage{tikz}
\documentclass[default]{sn-jnl}

\usepackage{amsmath,amssymb,amsfonts}
\usepackage{graphicx}
\usepackage{textcomp}
\usepackage{subcaption}
\usepackage{booktabs}
\usepackage{adjustbox}
\usepackage{array}
\usetikzlibrary{arrows, shapes}
\tikzstyle{block} = [rectangle, draw, fill=black!20,  text centered, rounded corners, minimum height=3em]
\usepackage{color,listings}

\jyear{2022}%

\theoremstyle{thmstyleone}%
\theoremstyle{thmstyletwo}%

\theoremstyle{thmstylethree}%

\begin{document}

\title[ ]{Explainable AI Algorithms for Vibration Data-based Fault Detection: Use Case-adadpted Methods and Critical Evaluation}

\author*[1]{\sur{Oliver Mey}}\email{oliver.mey@eas.iis.fraunhofer.de}

\author[2]{\sur{Deniz Neufeld}}\email{deniz.neufeld@uni-bamberg.de}

\affil[1]{\orgname{Fraunhofer IIS/EAS, Fraunhofer Institute for Integrated Circuits, Division Engineering of Adaptive Systems}, \city{Dresden}, \state{Germany}}

\affil[2]{\orgdiv{Cognitive Systems Group}, \orgname{University of Bamberg}, \city{Bamberg}, \state{Germany}}

\abstract{Analyzing vibration data using deep neural network algorithms is an effective way to detect damages in rotating machinery at an early stage. However, the black-box approach of these methods often does not provide a satisfactory solution because the cause of classifications is not comprehensible to humans. Therefore, this work investigates the application of explainable AI (XAI) algorithms to convolutional neural networks for vibration-based condition monitoring. For this, various XAI algorithms are applied to classifications based on the Fourier transform as well as the order analysis of the vibration signal. The results are visualized as a function of the revolutions per minute (RPM), in the shape of frequency-RPM maps and order-RPM maps. This allows to assess the saliency given to features which depend on the rotation speed and those with constant frequency. To compare the explanatory power of the XAI methods, investigations are first carried out with a synthetic data set with known class-specific characteristics. Then a real-world data set for vibration-based imbalance classification on an electric motor, which runs at a broad range of rotation speeds, is used. A special focus is put on the consistency for variable periodicity of the data, which translates to a varying rotation speed of a real-world machine. This work aims to show the different strengths and weaknesses of the methods for this use case: GradCAM, LRP and LIME with a new perturbation strategy.}

\keywords{condition monitoring, explainable AI, fault detection, machine learning, order analysis, vibration analysis}



\maketitle

\section{Introduction}
\label{sec:introduction}
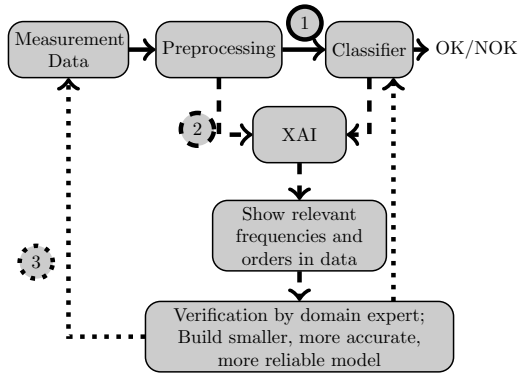
\begin{figure}[tpb]
	\centering
	\resizebox{0.6\linewidth}{!}{
		
		\begin{tikzpicture}[scale=0.5, every node/.style={scale=0.5}]
		\node  [block, text width=20mm] (v1) at (-6.4,7) {Measurement Data};
		\node  [block] (v2) at (-3.6,7) {Preprocessing};
		\node  [block] (v3) at (-0.8,7) {Classifier};
		\node   (v4) at (1.2,7) {OK/NOK};
		\node  [block,text width=15mm] (v6) at (-2.1,5.4) {XAI};
		\node  [block, text width=30mm] (v5) at (-2.1,3.5) {Show relevant frequencies and 
			orders in data};
		\node [block, text width=55mm] (v7) at (-2.1,1.6) {Verification by domain expert;\\
			Build smaller, more accurate,\\ more reliable model};
		
		\draw [->, very thick]  (v1) edge (v2);
		\draw [->, very thick, dashed] (v2) |- (v6);
		\draw [->, very thick, dashed] (v3)  |- (v6);
		\draw [->, very thick] (v2) edge (v3);
		\draw [->, very thick] (v3) edge (v4);
		\draw [->, very thick, dashed] (v6) edge (v5);
		\draw [->, very thick, dashed] (v5) edge (v7);
		
		\draw[<-, very thick, dotted](v3.310)--( v3.310 |- v7.north);
		\draw[->, very thick, dotted ]  (v7) -| (v1);
		\node[circle, draw, very thick, fill=black!20] (v9) at (-2,7.5) {1};
		\node[circle, draw, very thick, fill=black!20, dotted] (v8) at (-7,3) {3};
		\node[circle, draw, very thick, fill=black!20, dashed] at (-4,5.5) {2};
		\end{tikzpicture}
	}
	\caption{After the training in step (1), in step (2) XAI methods are used to generate saliency maps for the explanation of features. In step (3) the explanation is shown to domain experts to reduce the number of input features for future models.\label{fig:data_flow}}
\end{figure}

Rotating machinery finds application in many domains, from production facilities, wind turbines, automotive to aerospace and many more. 
Degradation and aging of machine components is an always ongoing process. Occurrence of failures is only a matter of time and determined by the runtime and load of a machine. However, once a defect develops in one of the components, operation will not run smoothly anymore, and the degradation rate of other components can also speed up due to increased vibrational load. Therefore, it is important to detect and eliminate defects as early as possible \cite{hashemian2011,nguyen2019,zhang2019}.\\
Faults in rotating machinery can lead to periodically occurring signatures in the vibration data of a system thereby enabling non-invasive and real-time condition monitoring of the state of a machine \cite{renwick1985,carden2004,vishwa2016,janssens2016}.
In this work, vibrational data is transformed to a frequency-RPM and an order-RPM map. In general, lines in the spectrum of the frequency-RPM map that are perpendicular to the time axis represent system specific frequencies which are excited by the resonance of the system. Lines whose position scales with the rotational speed of the system stem from rotating components like ball bearings or gear boxes. In the order-RPM map the x-axis is normalized based on the systems rotational speed. Vibrations caused by rotating elements are then shown as straight lines, while system resonance vibrations are sloped \cite{Swanson2005,Brandt2011}. State and defect classification algorithms can utilize those representations of the systems vibrations \cite{kateris2014,wang2019c}. In principle, this is already possible using hand-designed features and threshold values \cite{mcinerny2003,randall2011}. However, finding these features takes a lot of effort and requires expert knowledge of the inner workings of the machine \cite{sun2018}.\\
Deep learning algorithms on the other hand have the advantage that they require less effort in terms of data preprocessing and expert knowledge of the specific machine \cite{liu2016,liu2018, zhao2019,serin2020,zhang2020,mey2020}. However, deep learning necessitates a large enough amount of training data for each defect type or state. The collection of training data with defect information is often challenging and expensive. In addition, the vibration behavior of the same machine can differ significantly when the torque of the screws fixing the bearings changes or components unrelated to specific faults are exchanged \cite{mey2021}. However, if no change at the machine is conducted, the variability of the collected vibration data is low and measuring for longer amounts of time will not provide much more information to a data-driven model. Models are therefore usually prone to overfitting and have a high chance of not providing accurate classifications when applied to another machine identical in construction. Understanding the classification process inside a trained fault detection model is therefore crucial to identify cases where predictions are based on incorrectly learned input-output relationships and are thus not robust against changing operating conditions.\\
The classification process of large deep learning models is usually not comprehensible to humans. Overfitting of a model or classifications based on wrong reasons can stay hidden until the model is applied to new data. Explainable AI (XAI) algorithms aim to overcome this obstacle by making the classification process transparent \cite{adadi2018,guidotti2019,jeyakumar2020}. Many XAI algorithms were initially developed for applications in image recognition \cite{zhang2018}. Examples are GradCAM (Gradient-weighted Class Activation Mapping) \cite{selvaraju2017}, LRP (Layer-wise Relevance Propagation) \cite{bach2015}, LIME \cite{ribeiro2016}, DeepLift \cite{shrikumar2017}, Integrated Gradients \cite{sundara2017}, or SHAP (Shapley Additive Explanations) \cite{lundberg2017}. Applied to image recognition, the goal of XAI is to highlight areas of classified images, which influenced the model's decision for a certain class to a high extend. These visualizations are also called saliency maps. To apply these algorithms to condition monitoring, they have to be transferred to the case of one-dimensional data, i.e. time series. Still, it needs to be considered that saliency methods in some cases produce inconsistent results as reported by Kindermans et al. \cite{kindermans2019}. Since machine diagnosis can apply to safety critical domains, it is important to verify all algorithms used as well as possible.\\
Artificial intelligence and deep learning is a relevant topic in the domain of rotating machinery diagnosis \cite{nath2021}. First demonstrations of the transfer of XAI techniques to machine fault diagnosis have already been reported. Chen et al. used GradCAM to explain fault classifications of a convolutional neural network, which are based on short time Fourier transformed vibration data \cite{chen2020}. GradCAM was further applied to explain bearing fault classifications based on preprocessed acoustic emission data \cite{kim2020}, vibration data based classifications of a neuro-fuzzy network \cite{lin2021} and in \cite{saeki2019} GradCAM was applied in an anomaly detection use-case for time series of vibration data of a rotating system. GradCAM was also applied to explain vibration data-based fault detection of linear motion guides \cite{kim2021}, bearings \cite{yoo2022} and grinding machines \cite{liu2022}. Further, frequency activation maps were calculated to explain a time domain-based bearing fault detection model in frequency domain \cite{kim2021b}. LRP was used to explain fault classifications based on vibration data from a gearbox \cite{grezmak2019} and of multi-sensor information from a motor \cite{grezmak2020}. SHAP was used in \cite{hasan2021} to explain bearing fault classifications of a k-nearest neighbor classifier. The attention mechanism was utilized to explain bearing fault classifications from time domain vibrational data in \cite{li2019} and \cite{wang2020b}.\\
For image classification tasks the comparison of such an analysis with a ground truth is easily possible for humans as the ground truth is just the area of the object to be classified. However, the comparison with a ground truth for the case of vibrational data is not obvious. A comparison of different XAI algorithms for measurement data is therefore challenging. \\
This work aims to evaluate the plausibility of XAI methods explaining deep neural networks applied to vibration-based condition monitoring with a special focus on machines with variable rotation speed. We therefore created a synthetic data set based on sinusoidal data, allowing for an intuitive comparison of XAI algorithms applied to a supervised classification task based on this data set to a ground truth. Visualized as frequency-RPM map as well as an order-RPM map, rotation speed-dependent and -independent modes can be separated visually, enabling a use-case specific evaluation of various XAI outputs. The three investigated XAI algorithms GradCAM, LRP, and a modified version of LIME are further applied to an imbalance detection data set \cite{fordatis2020} in the following to transfer the findings from the synthetic data set to real-world vibrational data. The obtained additional information from the classifications can be used, for example, to restrict the input of machine learning models to certain frequency ranges as shown in Figure \ref{fig:data_flow} and to support domain experts verify the predictions of CNNs for vibration analysis. As a side result, the highest classification accuracy on the imbalance data set reported so far was improved to $99.66 \%$. To increase reproducibility, the source code of all analyses performed is published at Github \cite{our_github}. In the Supporting Information published along with this paper, the studies shown here are supplemented with those using additional XAI methods.
\section{Methods}
\label{sec_methods}
In the following, the data sets used in this work are described. Then a more detailed explanation of the preprocessing steps taken is shown before the classifier models and the XAI approaches are explained. 
\subsection{Data Sets}
Two different data sets are demonstrated. One is based on a superposition of sinusoidal functions and is designed as part of this work to serve as an evaluation tool of the investigated XAI methods. The second is an actual vibrational data set recorded from a rotating machine. 
\subsubsection{Sine Cut-off Classification}
\begin{figure}[tpb]
	\centering
	\includegraphics[width=\linewidth]{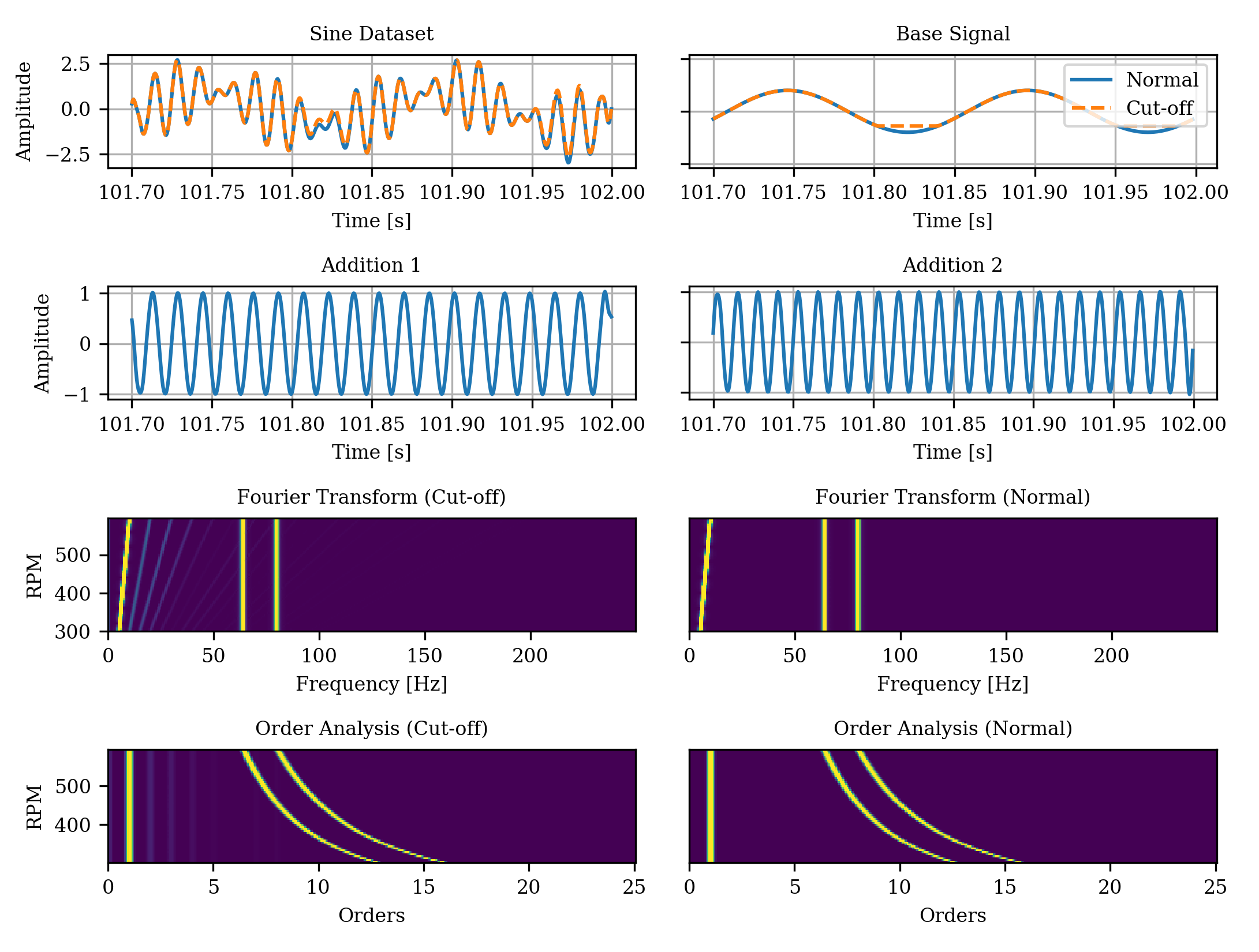}
	\caption{Sine data set components. Either the "Normal" or "Cut-Off" data is used and "Addition 1" and "Addition 2" are added.}
	\label{fig:sine}
\end{figure}

In typical XAI and saliency map applications, the output for the classified image is intuitively verifiable and usually corresponds to the object in the image or characteristic features of it. Vibration data is only partially intuitively understandable and needs deep domain knowledge on the underlying system. However, a visual comparison of different XAI algorithms is difficult without intuitive input data. We therefore introduce a synthetic data set for a binary classification of a periodic time series. This consists of additively combined sine functions, as shown in Figure~\ref{fig:sine}. The main component is a chirp (sine with linearly increasing frequency), which is truncated at values of amplitude $<-0.7$ to $0.7$ for one class which is referred to be the \textit{cut-off} class. The class without truncation is referred to be the \textit{normal} class. The two other components are two sine functions which have two different, higher and constant frequencies. They are added to both the normal and the cut-off sequences. In the time domain, which is however not considered in this paper, the truncated regions are the distinguishing feature and should be highlighted by an XAI algorithm. In the frequency domain, and thus in the order domain as well, the cut-off causes a perturbation that produces amplitudes at orders higher than the fundamental mode. Accordingly, these should also be highlighted by an XAI algorithm as a distinguishing feature. 
\begin{figure}[tpb]
	\centering
	\includegraphics[width=\linewidth]{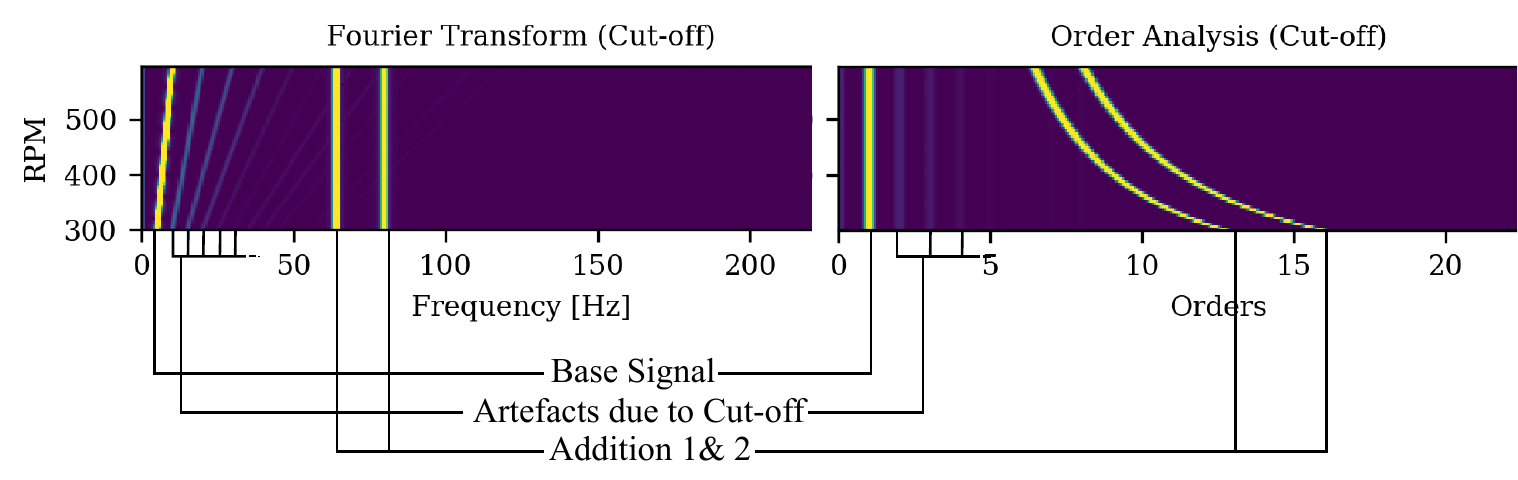}
	\caption{Elements within the spectrum of the Sine cut-off dataset in the frequency-RPM map and the order-RPM map representation.}
	\label{fig:todo}
\end{figure}
\subsubsection{Imbalance Classification}
The data set used for the demonstration of the methodology in a real-world application was introduced in \cite{mey2020}. It contains measurements obtained from vibration sensors attached to a drive train of a rotating motor, where imbalances of various strength were mounted. There are two measurements for each imbalance strength, one for model training and one for validation, respectively. During each measurement, the rotation speed of the drive train was gradually increased from a lower motor RPM 
to a higher RPM for two times. It thereby allows for a rotation-speed dependent evaluation of the classification accuracy of imbalance detection models. The data set is available at \cite{fordatis2020}.

\subsection{Preprocessing}
Changes in a rotating system -- like defects in its ball bearings or an evolving imbalance -- causes distinct changes in the Fourier spectrum of its vibrational signal \cite{Brandt2011}. 
Here, we conducted both a transformation into a frequency-space representation as well as into an order space representation to be utilized as classification input.\\
For the frequency space representation we applied a frequency-RPM transformation. It is defined as a repeated application of the Fast Fourier Transform (FFT) to a small temporal subsection of the complete signal at increasing rotational speeds, similar to a Short-time Fourier transform (STFT). This yields a better temporal resolution of the FFT per time interval compared to the application of an FFT over the complete time series. With rising RPM, as simulated in the sine cut-off data set and given in the imbalance data, resonance frequencies characteristic to the system appear in the corresponding frequency-RPM map as lines parallel to the time axis. Conversely, lines in the map that belong to resonance frequencies that stem from rotating components like motors, pumps or ball bearings, increase linearly with rising RPM of the system \cite{Brandt2011}.\\
The order-RPM transformation applied as well is similar to the frequency-RPM transformation with the difference that here, the transformed signal is interpolated to a representation normalized to the current rotation speed of the system. Visualized into an order-RPM map, the x-axis of this plot would be given in orders with an order of 1 corresponding to the current RPM value. This way, the correlation of spectrum lines inverses: Lines parallel to the time axis now belong to rotating subcomponents. Oscillations of constant frequency on the other hand will appear as curved lines. The explained transformations were performed in Matlab using the \lstinline$rpmfreqmap$ and \lstinline$rpmordermap$ methods.\\
The background knowledge on these transformations also explains the usage of our sine cut-off data set, since its spectrogram shows similar features like actual vibrational data from drive trains.\\
The resolution of the frequency-RPM representation was chosen such that the resulting map has a satisfactory temporal and frequential resolution to the human eye when visualized accordingly. To make the classification results comparable, the resolution of the order-RPM representation was chosen such that the resulting feature vectors have roughly as many elements as the feature vectors of the frequency-RPM representation.
\subsection{Classification Algorithms}
\begin{figure}[tpb]
\centering
\includegraphics[width=0.5\linewidth]{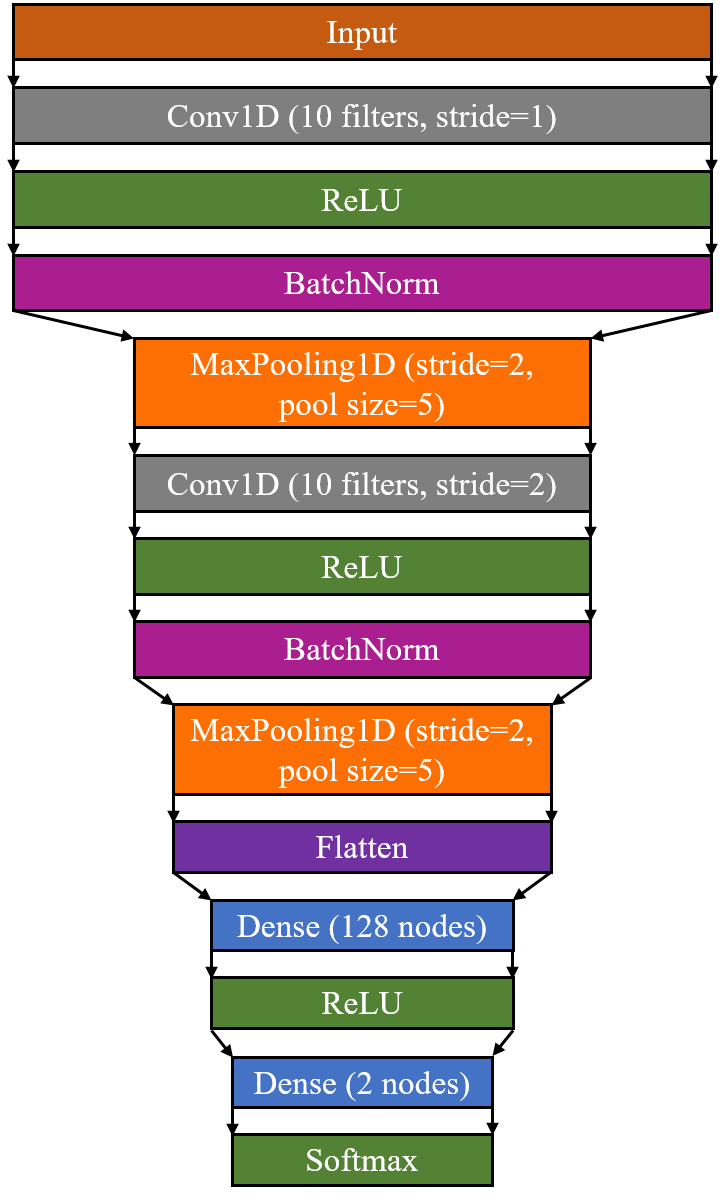}
\caption{Layer-by-layer neural network architecture of the used classification models. While the general architecture is identical for the considered classification tasks, the dimension of each layer differs due to the task-specific dimension of the input vector.}
\label{fig:nn_architecture_general}
\end{figure}

The classification of the data sets was performed using CNNs. This allows the application not only of model-agnostic XAI methods like LIME but also model-specific ones like GradCAM or LRP. The layer-by-layer network architectures used are visualized in Figure \ref{fig:nn_architecture_general}.
In each case, two convolutional blocks consisting of a convolutional layer with a ReLU activation function followed by batch normalization and a MaxPooling layer are used. Then, a flattening layer is applied and, using two dense layers with again ReLU activation, the final classification is provided using two output nodes with a softmax activation function. The models were each trained for 150 epochs using the Adam optimizer with a learning rate of $1e-4$. A model checkpoint was used to store the model weights with the best prediction accuracy on the test data set during the optimization process.
\subsection{XAI algorithms}
\label{sec:methods_xai_algorithms}
\begin{figure}
	\centering
	\includegraphics[width=0.9\linewidth]{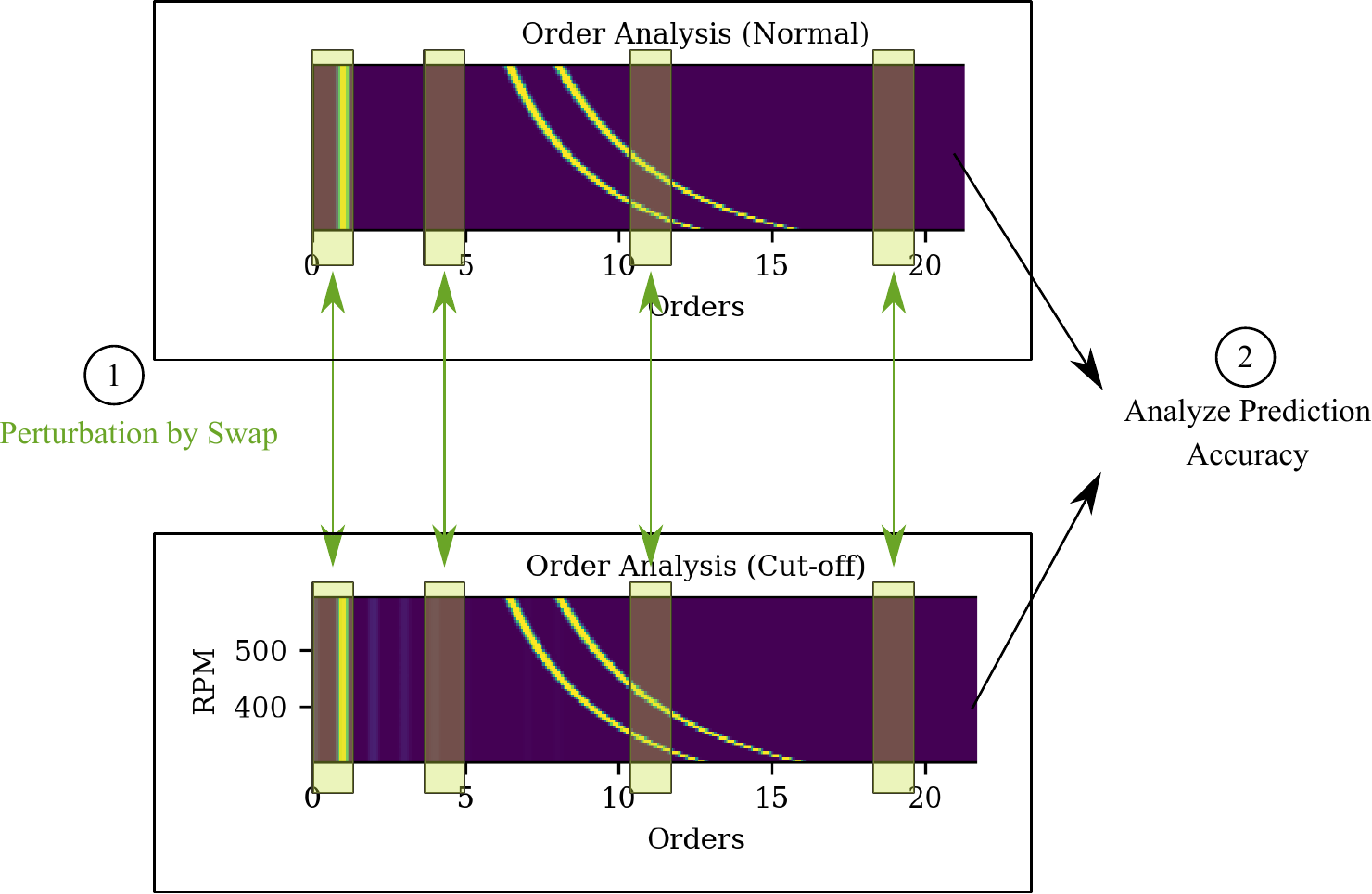}
	\caption{Schematic on the proposed perturbation strategy. For the input data of each class, data in the segment is replaced by data from the opposite class at the same segment position in the spectrum.}
	\label{fig:limeglobal}
\end{figure}

The XAI algorithms used in this work are GradCAM, LRP-Z and LIME. Additionally, the Supporting Information contains similar investigations for GradCAM++, ScoreCAM, LRP-$\varepsilon$ and another LIME implementation for time series as described in the following. They aim to highlight, which areas of a specific input sample influenced the model output most.\\
Class Activation Maps (CAM) \cite{zhou2016} can be extracted by calculating the weighted sum of the feature maps of the last convolutional layer. However, for CAM it is required that a global average pooling layer and the final output layer directly follow the last convolutional layer. This obstacle was overcome by the introduction of GradCAM \cite{selvaraju2017}, where the gradients between the last convolutional layer and the output nodes are used to weight the individual filter channels for the calculation of the overall activation map. GradCAM++ \cite{chattopadhay2017} modifies the calculation of the gradient-based weights to achieve better results for cases, where multiple instances of a certain class appear in a single data sample. Since both GradCAM and GradCAM++ rely on the gradients to weight the feature maps, no activation map can be extracted for cases, where no gradient exist, i.e. super-confident predictions. To remedy those, label smoothing was applied to the cross-entropy loss used for model optimization. This means that the target values for a specific class are not 0 or 1 anymore but in our case 0.05 and 0.95. Consequently, it is possible to extract a gradient from the final softmax layer also for the case of the numerically required finite precision. ScoreCAM \cite{wang2020} overcomes the super-confidency issue by obtaining feature map weights by calculating the channel-wise increase of confidence of the model, when the original input is multiplied with the upsampled feature maps. ScoreCAM does therefore not depend on the model gradients. Further the authors showed that it localizes class-discriminant input areas more accurate compared to GradCAM and GradCAM++.\\
LRP (Layer-wise relevance propagation) by Bach et al.\cite{bach2015} is a method aimed at product type nonlinear networks like convolutional neural networks. It provides a relevance score per pixel based on a classifier and an input image. Backtracking from the output layer of the network, this method computes the influence on the result based on the prior layer, layer for layer, until the input of the network. This way, the importance of each input pixel is determined. The basic version of LRP is called LRP-Z. There are several modifications of LRP such as LRP-$\epsilon$, while LRP-Z based on the implementation of ref. \cite{alber2019} was selected due to the clearer saliency maps resulting from the conducted investigation. \\
Local Interpretable Model-Agnostic Explanations (LIME) by Ribeiro et al. \cite{ribeiro2016} is a model-agnostic method for quantifying feature importance. It perturbs the input data instance several times in random places by replacing parts of it with e.g. the data set's average, noise, or zeros. After applying the trained model to the new data again, a ridge regression model is used to find the perturbations that influence the model's outcome the most. This way, the important parts of the original instance can be found. LIME has been applied in related work to tabular, text, time series and image data. \\
The implementation of our LIME variant was developed based on a version of LIME for time series described and available in \cite{lime4time}. In that source, an input sequence is split in a preset amount of segments and perturbed as described before. LIME can be computationally expensive for large data sets, because it's typically applied per data set instance. Due to our large amount of data, this would not be effective. But the fact that certain frequency and order bands are important for the anomaly classification of rotating systems can be leveraged in our case. To overcome the limitation due to large data, instead we propose a new perturbation strategy for LIME (Global LIME) for global feature importance for RPM-maps by applying LIME to the complete input data set at once. In the frequency domain, this means that certain frequency bands are perturbed, and the same goes analogously for the order analysis. To generate examples of better fidelity, instead of with noise or zeros we replace the bands in one class with the data from the same position in the opposite class (as shown in Figure \ref{fig:limeglobal}). Other replacement strategies are investigated in the Supporting Information. In the original Lime-for-time implementation, the amount of segments $n$ and the amount of features $f$ act as hyperparameters. In our experiments, it became apparent that the choice of these parameters had a large impact on the analysis results, making the method less robust. To achieve more consistent results than with just one configuration, we do not just choose one combination of $n$ and one $f$. Instead, we generate LIME evaluations for multiple segment counts $n_i$ and feature counts $m_j$ for all replacement strategies. This way, $i*j$ intermediate output saliency maps are created. The final saliency map is computed by averaging all resulting maps using their mean. 
\subsection{Visualization}
The goal of XAI is to provide humans with insights into how black-box machine learning algorithms work. A crucial aspect in the use of XAI methods is therefore the processing of the additional information obtained. In this work, the determined saliency values are visually processed in frequency-RPM maps and order-RPM maps. Using the Python library Matplotlib, all diagrams were created as heatmaps with the Viridis \cite{viridis} color map. Its advantages are a perceptually uniform representation of colors and a good translation to grayscale, so it is very suitable even in case of color blindness. \\
Prior to plotting, the saliency values obtained as output from the applied XAI methods were preprocessed to maximize the visibility of the patterns they contain. All XAI output data was first normalized to a range between 0 and 1. For GradCAM and LRP-Z, only positive values were used to mitigate visual noise while this was not necessary for the case of LIME. GradCAM and LIME were then plotted with a linear color scaling with the upper boundary of the corresponding colormap based on the 0.95 quantile to compensate for outliers. LRP had to be visualized with logarithmic color scale to reduce image noise within the results, especially for the imbalance classification data set.
\section{Results}
The classification models were trained using their training and test data sets as described in Section 2.1. In the following, the validation accuracies achieved in each case are validated and subsequently the outputs of different XAI algorithms, which are supposed to explain the classifications of the models, are visually evaluated.
\subsection{Classification Accuracy}
The prediction accuracy of the models described in Section \ref{sec_methods} were evaluated after their training using the respective validation data sets. Results are shown in Tables \ref{tbl:xai_acc} and \ref{tbl:xai_loss}. In the case of the sine Cut-off data set, $100 \%$ classification accuracy was achieved both when using the Fourier transformed data and when using the order analysis. Apparently, the detection of the inserted disturbance in the sinusoidal function values is a comparatively easy task for CNNs. Validation accuracies of $99.66 \%$ and $98.49 \%$ were achieved for the classification of the imbalance data set, using Fourier transform and order analysis, respectively. Thus, despite identical model architecture, higher classification accuracy was achieved with the Fourier transformed data. Still, both classification accuracies are higher compared to the best value achieved so far, which was reported to be $98.2 \%$  \cite{mey2020}.
\begin{table}[htpb]
	\centering
	\caption{Classification loss on test set}
	\label{tbl:xai_acc}
	\begin{tabular}{@{}lll@{}}
		\toprule
		& \multicolumn{2}{c}{Domain} \\ \cmidrule{2-3}
		Data set   & Frequency & Order  \\ \midrule
		Sine Cut-Off &  5.92 & \textbf{5.26}  \\
		Imbalance &  \textbf{0.92} & 1.28  \\ \bottomrule
	\end{tabular}
\end{table}
\begin{table}[htpb]
	\centering
	\caption{Classification accuracy on the test set}
	\label{tbl:xai_loss}
	\begin{tabular}{@{}lll@{}}
		\toprule
		& \multicolumn{2}{c}{Domain} \\ \cmidrule{2-3}
		Data set    & Frequency & Order  \\ \midrule
		Sine Cut-Off &  \textbf{1.000} &\textbf{ 1.000}  \\
		Imbalance& \textbf{0.996} & 0.976  \\ \bottomrule
	\end{tabular}
\end{table}
\subsection{XAI evaluation: Sine Cut-off Classification}
\begin{figure*}[h!]
	\centering
	\begin{subfigure}[b]{.85\textwidth}
		\includegraphics[width=\linewidth]{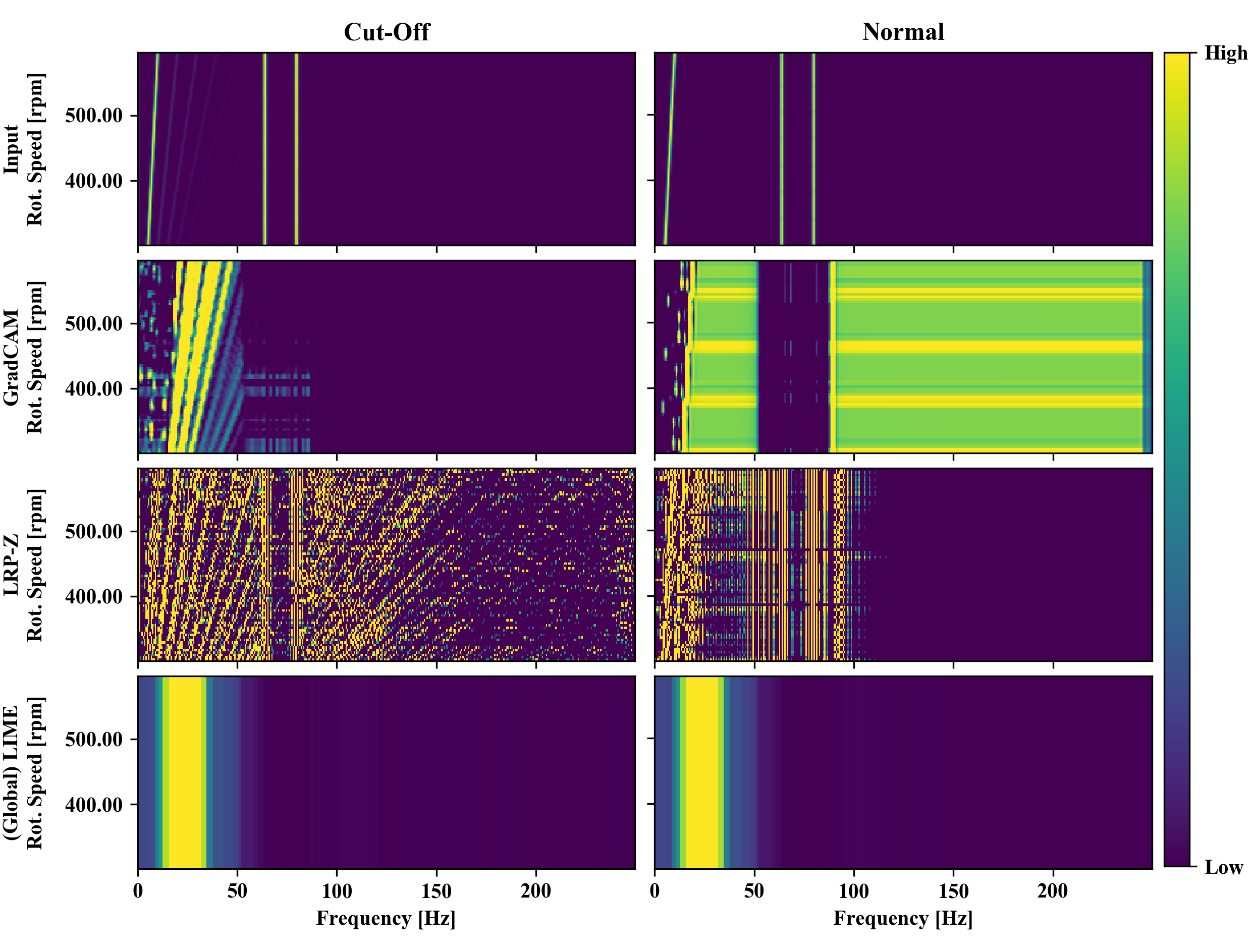}
		\caption{Frequency-RPM map}\label{fig:sine_fft_result}
	\end{subfigure}\\
	\begin{subfigure}[b]{.85\textwidth}
		\includegraphics[width=\linewidth]{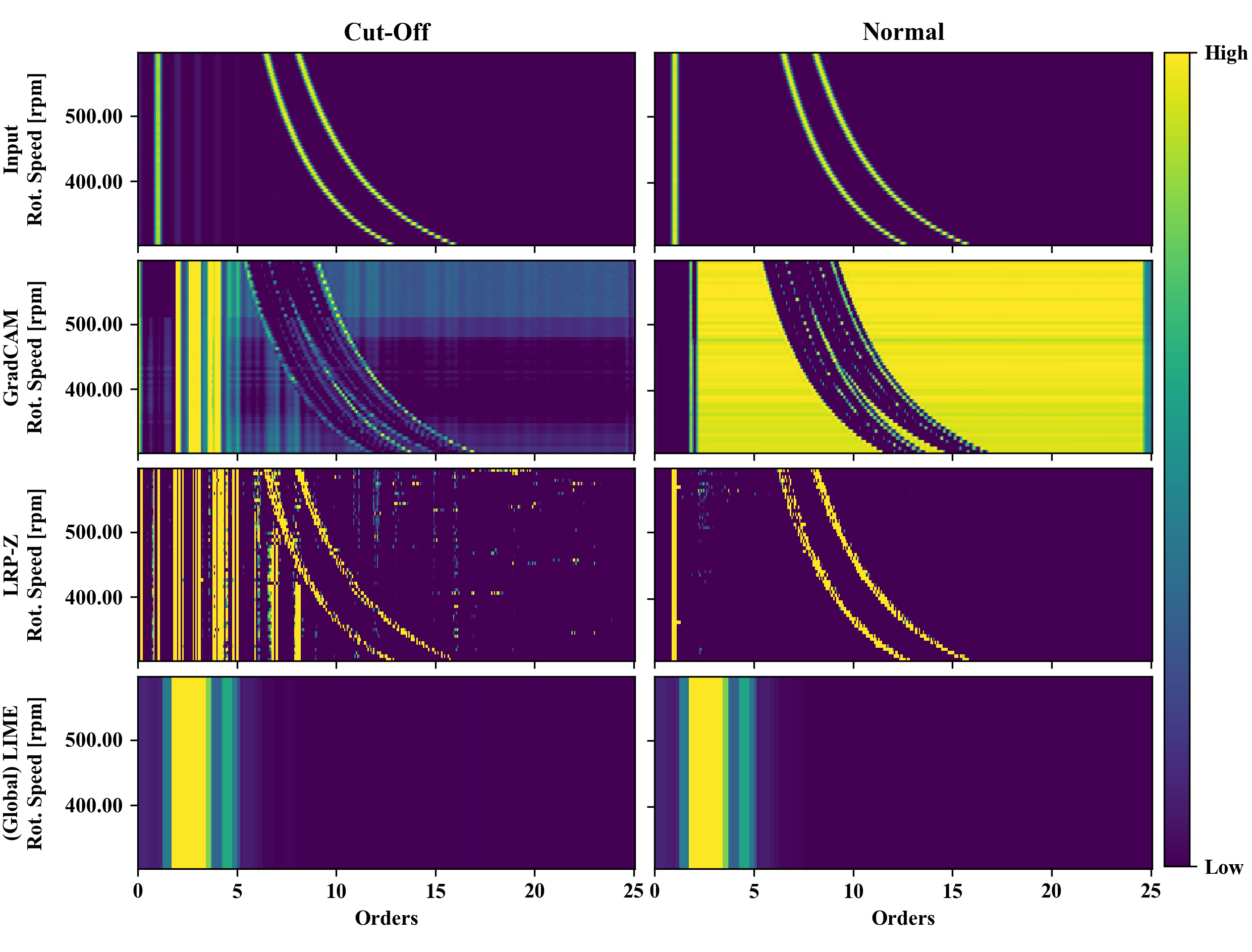}
		\caption{Order-RPM map}\label{fig:sine_order_result}
	\end{subfigure}
	\caption{Visualization of the input data used for model training as well as saliency maps of three XAI methods applied to sine cut-off classification in a frequency-RPM-map (a) and an order-RPM-map (b).}\label{fig:result_sine}
\end{figure*}

In the next step, various post-hoc XAI methods are applied to the trained classification models. Figure \ref{fig:sine_fft_result} shows the raw frequency-RPM maps of the  input values as well as the outputs of the XAI methods GradCAM, LRP-Z and (Global) LIME. The left column of the figure shows the data for the case with cut-off, the second column the data for the normal case (without cut-off).
For the case of the  visualization, the signals corresponding to the basic sine function with increasing frequency, in the following called base mode,  are seen as the line in spectrum closest to the rpm-axis.  The signals of the added sine functions with constant frequency (the two lines the farthest away from the rpm-axis, hereafter referred to as the superimposed modes) are also clearly visible. 
The distortion induced by the cut-off of the base signal mode is visible by the additional higher modes at multiples of the frequency of the fundamental mode. While the \textit{Cut-off} class is detectable at the higher orders of the fundamental mode, the \textit{normal} class is detectable by the absence of those higher orders as well as by a higher amplitude of the fundamental mode. Ideally, an XAI algorithm would ignore the superimposed modes as they provide no distinguishing feature between the two classes.\\
The output of GradCAM fulfills the condition, that it should ignore the superimposed modes. Additionally, the higher orders of the fundamental mode are highlighted for the \textit{cut-off} class. For the case of the class \textit{normal}, areas are highlighted which seem to contain no class-specific information, especially for the case of frequencies higher than those of the superimposed modes. \\
LRP-Z highlights even more higher orders of the fundamental mode when classifying the \textit{cut-off} class, while on the other hand also highlights the superimposed modes. Those superimposed modes, together with the fundamental mode, are also highlighted when classifying the \textit{normal} class. This can either happen because of the Log-scaling or the fact, that LRP is enhancing features of low amplitude. In any case, it provides a better contrast than the original presentation of the input, but no additional information on the functioning of the CNN.\\
The implementation of LIME for time series introduced in this paper provides global information about the relevance of individual features. Accordingly, in this case for the class \textit{normal} the lowest regions of the spectrum should be highlighted, which also contain the fundamental mode. For the class \textit{cut-off}, on the other hand, the areas between the fundamental mode and the superimposed modes should be highlighted. For both classes, the global LIME variant highlights the regions of the spectrum that lie directly above the fundamental mode, i.e., the regions where the distortion caused by the cut-off occurs. While this is not identical to the above conditions, the binary nature of the classification task implies that this is nonetheless a correct explanation. Accordingly, classification is performed by evaluating whether or not the higher orders of the fundamental mode are present.\\
In addition to the evaluation on the frequency-RPM map, the same procedure was also applied to the order-RPM map of the data (shown in Figure \ref{fig:sine_order_result}). The position of the fundamental mode is fixed to order 1 regardless of the rotational speed and therefore forms a vertical line in the diagram in the top row of Figure \ref{fig:sine_order_result}. The disturbance caused by the cut-off of the sinusoidal function can be seen at integer numbers of the corresponding order higher than 1. The superimposed modes, on the other hand, form a curved line in the order-RPM map. \\
The output of GradCAM is similar to its application to the frequency-RPM map: For the \textit{cut-off} class, the higher orders of the fundamental mode are highlighted and the superimposed modes are ignored. The unexpected highlighting of the higher order regions for the \textit{normal} class also appears similarly. \\
When LRP-Z is applied, both the fundamental mode and its higher orders and the superimposed modes are highlighted similar to the result in Figure \ref{fig:sine_fft_result}. Compared to the application on the frequency-RPM map, the fundamental mode is more emphasized in the \textit{normal} class as it would be expected from the design of the data set. \\
With Global LIME, the order range above the fundamental mode is highlighted for both classes, while the fundamental mode itself and the superimposed modes are not highlighted. This means that Global LIME correctly distinguishes between relevant and non-relevant features. However, due to the global nature of this explanatory method, a comparatively broad band is highlighted, since the entire frequency range within the sweep of the rotation speed needs to be covered.
In summary, for the synthetic data set considered here, Global LIME highlighted the relevant differences between the \textit{normal} class and the \textit{cut-off} class, but in a lower resolution than GradCAM and LRP-Z and without producing sample-specific explanations. GradCAM appeared to actually highlight important parts of the spectrum, but also highlighted non-relevant parts within the spectrum of the\textit{normal} class. LRP-Z visualized the input values with a modified intensity and thereby made features of a smaller amplitude more visible, but failed to mark the superimposed modes as non-relevant. 
\subsection{XAI evaluation: Imbalance Classification}
\begin{figure*}[h!]
	\centering
	\begin{subfigure}[b]{.85\textwidth}
		\includegraphics[width=\linewidth]{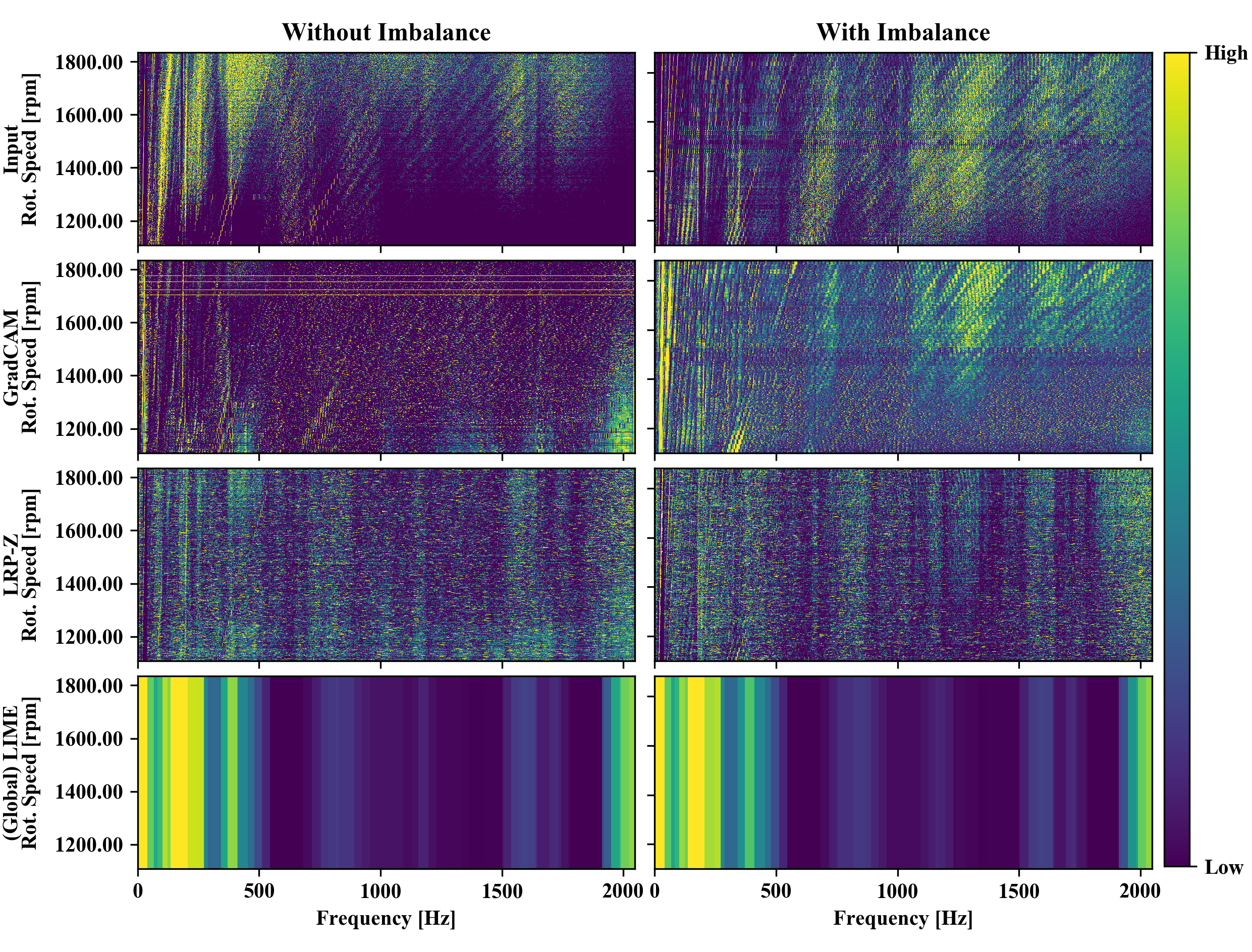}
		\caption{Frequency-RPM map}\label{fig:fhg_fft_result}
	\end{subfigure}\\
	\begin{subfigure}[b]{.85\textwidth}
		\includegraphics[width=\linewidth]{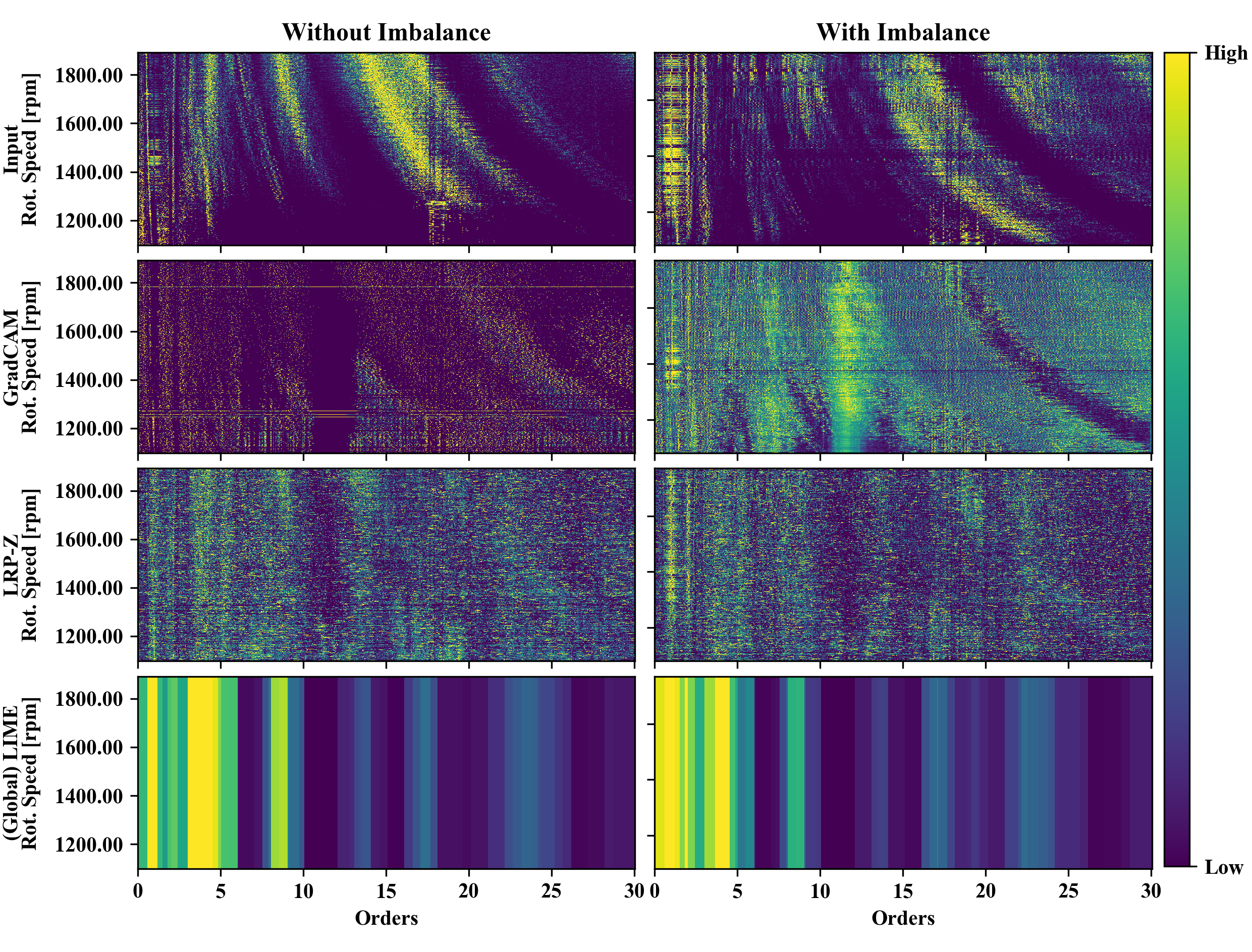}
		\caption{Order-RPM map}\label{fig:fhg_order_result}
	\end{subfigure}
	\caption{Visualization of the input data used for model training as well as saliency maps of three XAI methods applied to imbalance classification in a frequency-RPM-map (a) and an Order-RPM-map (b).}\label{fig:result_fhg}
\end{figure*}

The same XAI evaluation previously discussed for the sine cut-off classification was also performed for the imbalance classification. The corresponding heatmaps highlighting the input parts relevant for the classification according to different XAI algorithms are shown in Figure \ref{fig:result_fhg}.
The FFT-transformed vibration values are displayed in Figure \ref{fig:result_fhg}(a). Both, diagonal lines corresponding to vibrations with a frequency proportional to the rotational speed of the drive train as well as lines with a constant frequency corresponding to resonances inside the setup are clearly visible for the cases with and without imbalance.\\
The application of GradCAM to the resulting CNN classifier leads to a strong highlighting of the lower speed-dependent modes for the case of the imbalance class. This seems reasonable since an imbalance leads to oscillations with the periodicity of the rotational speed. For the imbalance free case a weak highlighting of the fundamental mode is present but not for its higher orders. The saliency map produced by GradCAM for the imbalance-free case on the other hand only highlights a small area with constant frequency at the upper  edge of the frequency range. For the imbalance case, many higher-order rotation speed-dependent modes are highlighted indicating that those modes are also excited by the imbalance. Low emphasis is given to the modes with constant frequency. This behavior is similar to the observations made by the application of GradCAM to the sine cut-off data set.\\
The response from LRP-Z, on the other hand, mainly shows modes with constant frequency. For both the classes with and without imbalance, almost no frequency dependence can be seen. Only for the case with imbalance, a weak emphasis on rotation speed-dependent modes is visible. Further, the difference in the visual appearance of the LRP-Z results of both classes is much weaker compared to GradCAM which makes it harder to interpret the classifier based on the LRP-Z result.\\
(Global)LIME highlights the frequency range between the fundamental mode, the lower range of the frequency spectrum (below 500~Hz). Similar to GradCAM, high importance is given to the highest obtained frequencies around 2000\,Hz as well. Some less important bands can be seen at around 800\,Hz, 1200\,Hz and 1600\,Hz.\\
The order analysis of the vibration data contains a range up to the 25th order, as depicted in Figure \ref{fig:fhg_order_result}. In the resulting diagrams, in particular  the modes of constant frequency stand out as curved lines, while the modes with a frequency proportional to the rotation speed can be seen here as vertical lines. In the case of the imbalance class, an increased intensity of the lower speed-dependent modes can be  seen clearly, analogous to the frequency representation.\\
In comparison to the Fourier transformed data, GradCAM highlights the rotation-speed dependent modes much less in this case, making the explanatory saliency less informative. The contrast between relevant and non-relevant input parts according to GradCAM is quite low while still some focus is given on rotation speed-dependent modes. While the GradCAM visualization seems to be less informative compared to the frequency representation, LRP-Z is able to clearly highlight the first and the second mode for the imbalance class. This is a significant improvement in comparison to its output applied to the Fourier transformed data. However, it would not be reasonable to use the LRP-Z result for dimensionality reduction, since it appears that no part of the input is declared non-relevant. For this purpose, the Global LIME result depicted in the last row of Figure \ref{fig:fhg_order_result} suited better, as there is a clear distinction between relevant and non-relevant parts of the data. Most relevant orders are under 10, but there are important bands at 14, 17 and around 24. Still, investigations would need to be carried out for this application to find out whether classification accuracy can still be maintained at a high level when removing all the input parts with zero feature importance.
\section{Discussion}
In general, frequency and order maps both are already a means to provide additional insights into vibration signals, especially for the case of a system with variable rotational speed. By applying XAI methods on vibration data-based fault classification, further information can be gained about the fault-specific relevance of vibrational modes in different spectral positions. Still, scaling the output of these algorithms is an important factor for their interpretability, which makes them dependent on additional parameters like thresholds and quantiles.\\
Due to the design of the sine cut-off data set, it was possible to assess the plausibility of various XAI algorithms applied to periodic time series. As a result, it was found that none of the applied XAI algorithms led to saliency maps which highlight class-specific features and omit non-distinguishing features as well as produced sample-specific explanations. Instead, these conditions were met partially by every method but never completely. It is also worth mentioning that the results from GradCAM and LRP-Z were complementary, which motivates the investigation of potential combinations of both methods. The utilization of an artificially created data set, such as the sine cut-off data set, for the visual evaluation of XAI algorithms can be a crucial tool to develop XAI algorithms specialized to be applied on periodic time series, or data with relevant information in the Frequency domain. Due to its design, the insights gained from the application of XAI methods on the sine cut-off data set can be transferred to other fault detection tasks encompassing periodic time series such as acoustic emission or motor voltages. On the other hand, the study using the sine cut-off data set showed that the information extracted should be taken with caution as the methods are able to correctly filter out non-relevant parts but on the other hand also highlight features which do not contain classification relevant information.\\
The investigations based on the imbalance data set demonstrated that the saliency values determined by means of XAI can shift in the frequency and order spectrum in proportion to the change in rotational speed. Still, a relatively huge share of the input data is marked as relevant for classification by the employed algorithms. Future work could focus on reducing this proportion by maintaining high saliency values only for those features that cannot be removed without decreasing classification accuracy. Shapley values \cite{lundberg2017} as well as explanations based on attention mechanisms could help to achieve this goal. 
\section{Conclusive Remarks}
In many application areas, the black-box nature of deep neural networks motivates the development of methods that make machine learning-based classifications explainable. In the field of machine fault diagnosis, XAI algorithms promise to make deep learning models more comprehensible and thus more robust and powerful. Following this motivation, we investigated the application of the popular XAI methods GradCAM, LRP and LIME on the classification of vibration data transformed by means of an FFT as well as by an order analysis. To evaluate the plausibility of XAI results especially for the case of machines with variable rotation speeds, we designed and introduced the synthetic \textit{sine cut-off data set} which revealed their inherent characteristics in terms of the extraction of classification-relevant input features. As a result, all methods were partially able to extract the class-distinguishing features while omitting those without class-specific information. Applying those XAI algorithms to imbalance detection in a real-world data set, rotation speed-dependent modes and constant frequency system resonances could be visually separated by means of a visualization with frequency-RPM and order-RPM maps. XAI outputs could be compared by considering their characteristics revealed by the evaluation with the sine cut-off data set. Our study motivates the further development of explainable machine learning methods applied to condition monitoring use-cases, while also putting emphasis on verifying these algorithms with more understandable data sets as a proof of concept first.

\bmhead{Supplementary information}
This article has accompanying supplementary information containing evaluations with further XAI methods.

\section*{Declarations}
\begin{itemize}
\item Code availability: This paper has accompanying source code available at \nolinkurl{https://github.com/o-mey/xai-vibration-fault-detection}.
\item Data Availability: This study utilized vibration measurement data publicly available and documented in the Fordatis repository, accessible at \nolinkurl{https://fordatis.fraunhofer.de/handle/fordatis/151.2}.
\item Authors' contributions: OM contributed the initial idea of utilizing XAI methods to examine variable-speed vibration data for fault classification. DN contributed the idea of enhancing the classification explainability by investigating data transformed by means of an order analysis and the idea of modifiing LIME to a global use case. DN contributed the first iteration of the sine data set, which was then further developed by OM. The setup of the classification algorithms was done by OM and DN. GradCAM, GradCAM++ and ScoreCAM investigation was conducted by OM while LRP-Z, LIME, Integrated Gradients, Deep Taylor and DeConv-Net investigation was conducted by DN. OM and DN contributed equally to the preparation of the manuscript.
\end{itemize}

\begin{appendices}

\section{Supporting Information}\label{SI}
\begin{figure}
	\centering
	\includegraphics[width=\linewidth]{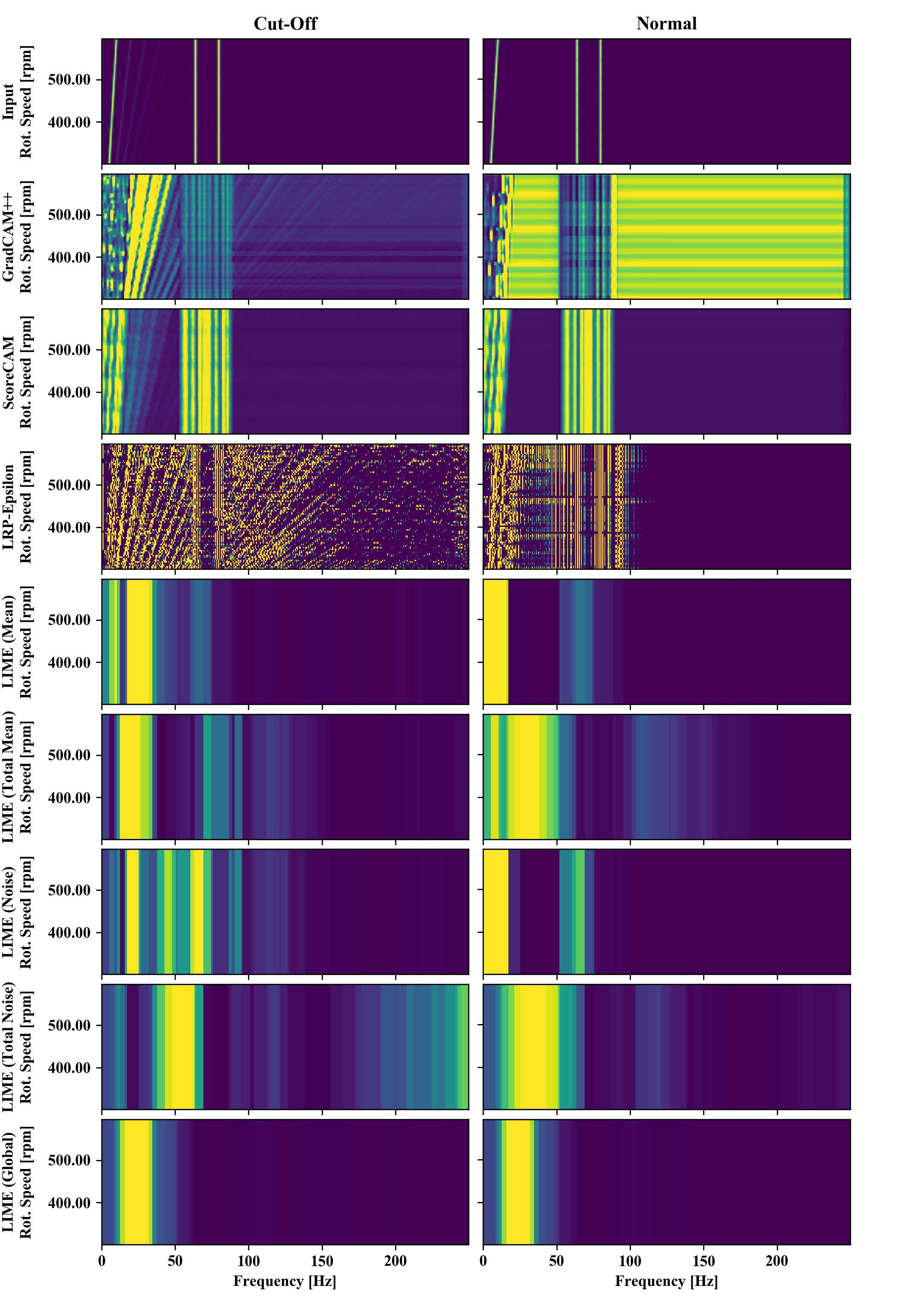}
	\caption{Visualization of saliency maps of the extended set of XAI methods applied to the sine cut-off classification in frequency representation.}
	\label{fig:sine_fft_si}
\end{figure}

\begin{figure}
	\centering
	\includegraphics[width=\linewidth]{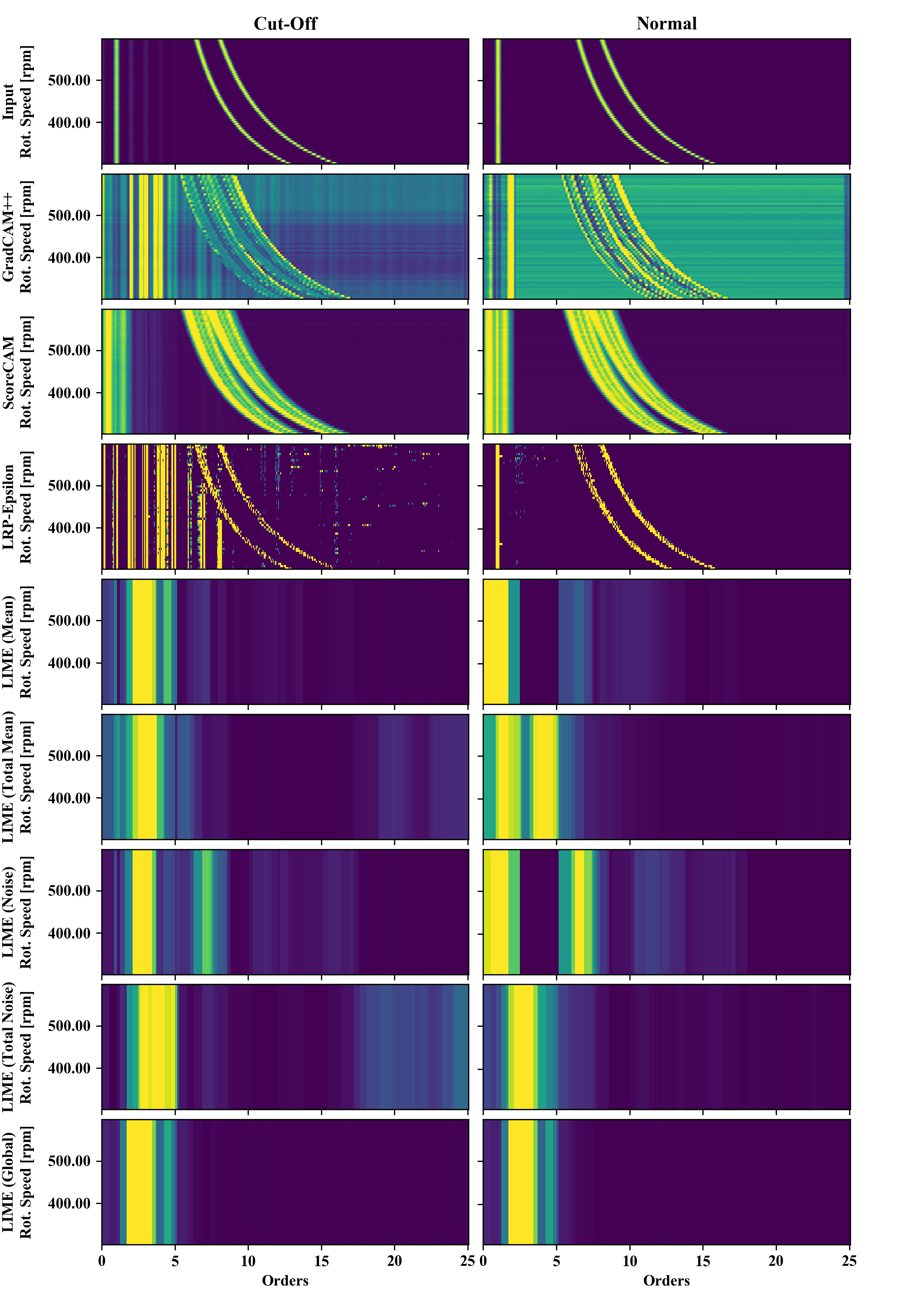}
	\caption{Visualization of saliency maps of the extended set of XAI methods applied to the sine cut-off classification in order representation.}
	\label{fig:sine_order_si}
\end{figure}

\begin{figure}
	\centering
	\includegraphics[width=\linewidth]{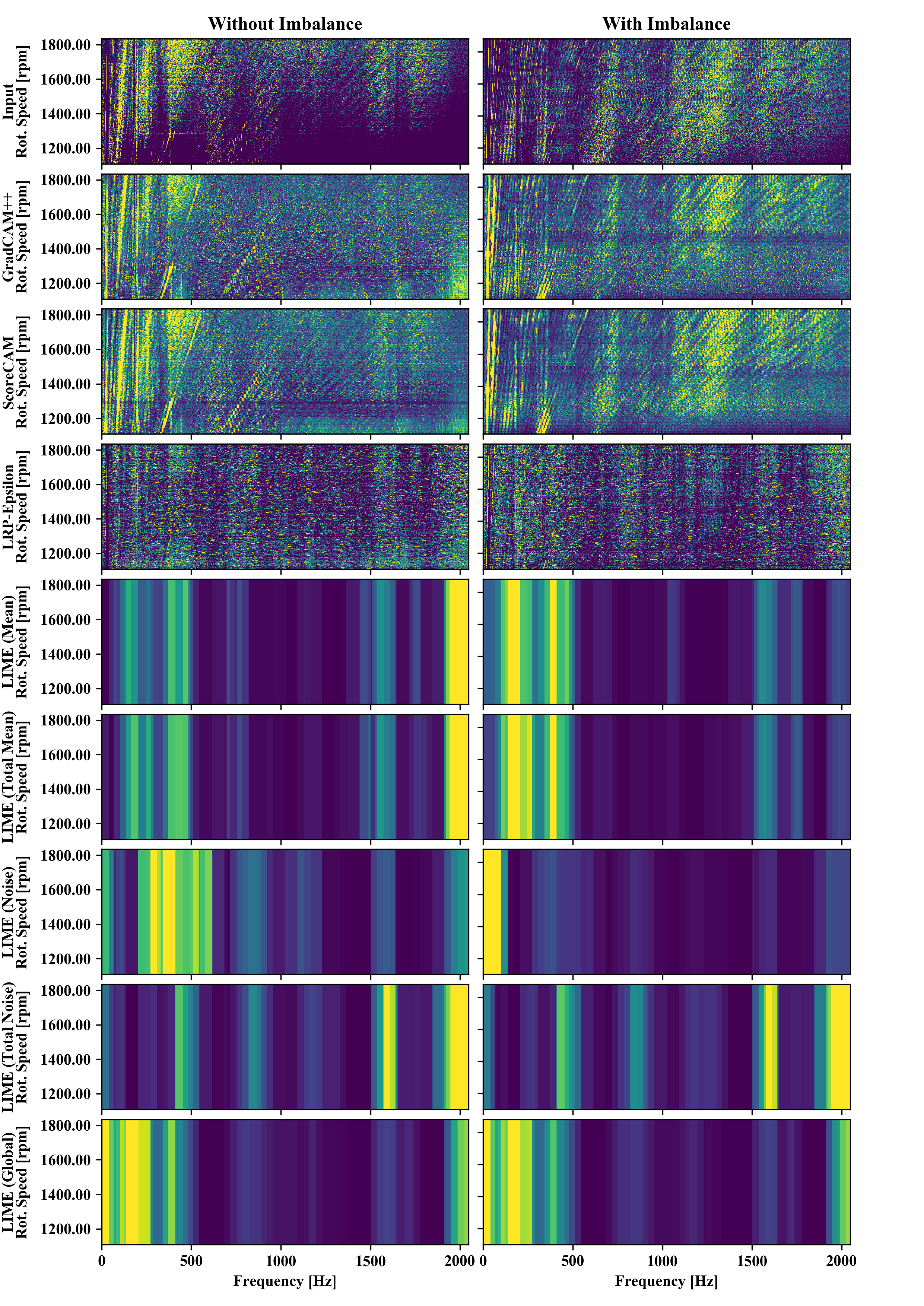}
	\caption{Visualization of saliency maps of the extended set of XAI methods applied to the imbalance classification in frequency representation.}
	\label{fig:fhg_fft_si}
\end{figure}

\begin{figure}
	\centering
	\includegraphics[width=\linewidth]{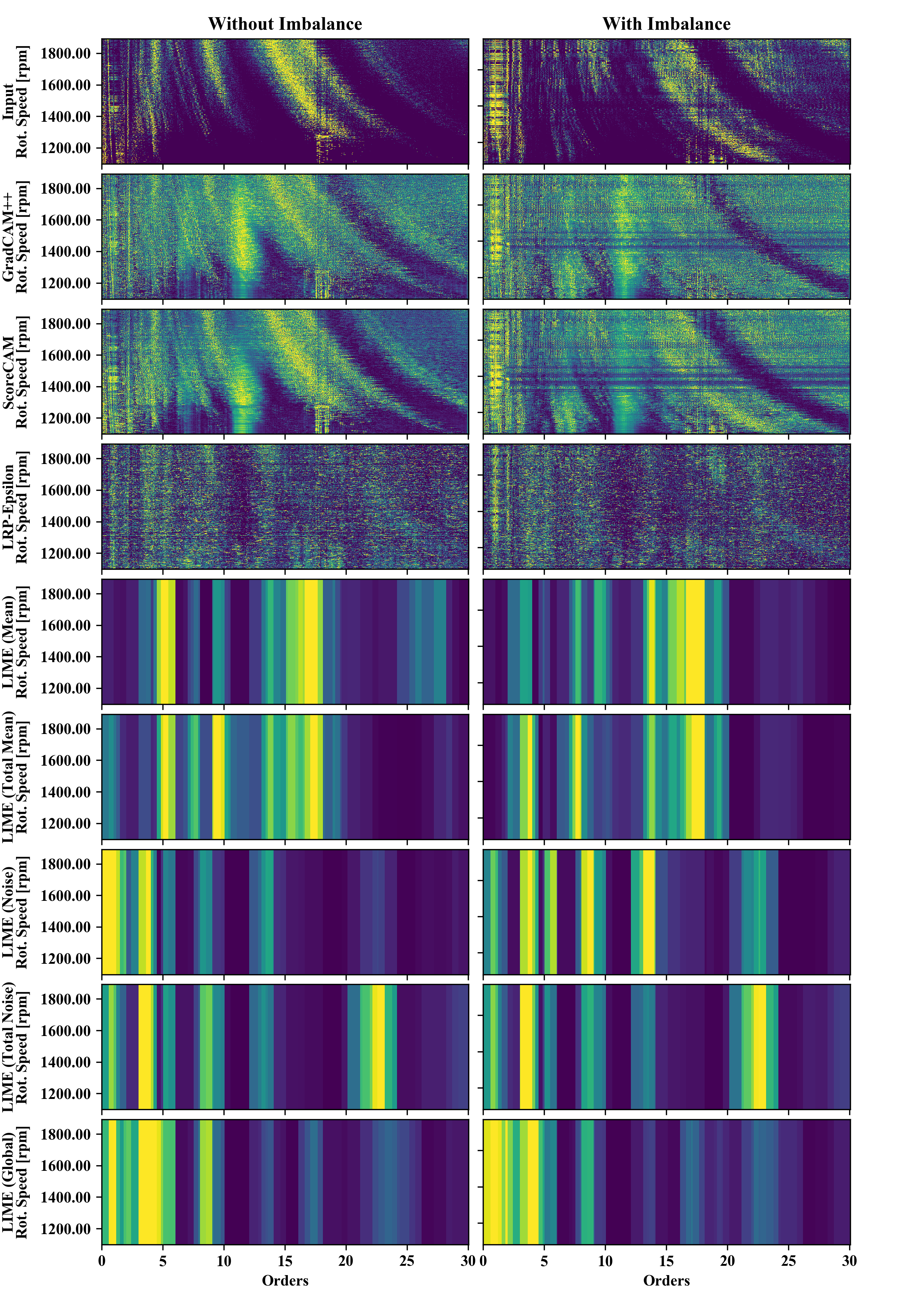}
	\caption{Visualization of saliency maps of the extended set of XAI methods applied to the imbalance classification in order representation.}
	\label{fig:fhg_order_si}
\end{figure}

In the following, the comparison of the different XAI algorithms is put on a broader basis. In addition to the methods GradCAM, LRP-Z, and the modified LIME implementation described in section \ref{sec:methods_xai_algorithms}, the methods GradCAM++, ScoreCAM, and LRP-Epsilon were also applied and examined. In addition, LIME was used similarly to the method described in the main section by replacing certain frequency bands and order bands; but instead of using values from the opposite class, noise with the same mean and standard deviation of the original data set was used.\\
The results of ScoreCAM and GradCAM++ were scaled in the same way as GradCAM from this paper, and LRP-Epsilon the same way as LRP-Z. The corresponding plots for the sine cut-off classification as well as for the imbalance classification for both the frequency and order representation are provided in the Figures \ref{fig:sine_fft_si}, \ref{fig:sine_order_si}, \ref{fig:fhg_fft_si} and \ref{fig:fhg_order_si}, respectively.\\
There are several different perturbation strategies reported for LIME. To give an impression of the performance of our method, some other standard perturbation strategies for LIME (in this case from \cite{lime4time}) are shown. The data are perturbed either with an average value or with uniform noise. The evaluated perturbation strategies are defined as follows:
\begin{itemize}
	\item Mean: Mean value of the  data from the original class at the perturbed sections
	\item Total Mean: Mean value of the complete data of the original class
	\item Noise: Noise with the value range from the data of the original class at the perturbed sections
	\item Total Noise: Noise with the value range from the complete data of the original class
\end{itemize}
\end{appendices}

\begin{thebibliography}{00}
\bibitem{hashemian2011} Hashemian, H. M. State-of-the-Art Predictive Maintenance Techniques. \textit{IEEE Trans. Instrum. Meas.} \textbf{60}, 226–236; 10.1109/TIM.2010.2047662 (2011).
\bibitem{nguyen2019} Nguyen, K. T. \& Medjaher, K. A new dynamic predictive maintenance framework using deep learning for failure prognostics. \textit{Reliability Engineering \& System Safety} \textbf{188}, 251–262; 10.1016/j.ress.2019.03.018 (2019).
\bibitem{zhang2019} Zhang, W., Yang, D. \& Wang, H. Data-Driven Methods for Predictive Maintenance of Industrial Equipment: A Survey. \textit{IEEE Systems Journal} \textbf{13}, 2213–2227; 10.1109/JSYST.2019.2905565 (2019).
\bibitem{renwick1985} Renwick, J. T. \& Babson, P. E. Vibration Analysis---A Proven Technique as a Predictive Maintenance Tool. \textit{IEEE Trans. on Ind. Applicat.} \textbf{IA-21}, 324–332; 10.1109/TIA.1985.349652 (1985).
\bibitem{carden2004} Carden, E. P. \& Fanning, P. Vibration Based Condition Monitoring: A Review. \textit{Structural Health Monitoring} \textbf{3}, 355–377; 10.1177/1475921704047500 (2004).
\bibitem{vishwa2016} Vishwakarma, M., Purohit, R., Harshlata, V. \& Rajput, P. Vibration Analysis \& Condition Monitoring for Rotating Machines: A Review. \textit{Materials Today: Proceedings} \textbf{4}, 2659–2664; 10.1016/j.matpr.2017.02.140 (2017).
\bibitem{janssens2016} Janssens, O. \textit{et al.} Convolutional Neural Network Based Fault Detection for Rotating Machinery. \textit{Journal of Sound and Vibration} \textbf{377}, 331–345; 10.1016/j.jsv.2016.05.027 (2016).
\bibitem{Swanson2005} Swanson, E., Powell, C. D. \& Weissman, S. A practical review of rotating machinery critical speeds and modes. \textit{Sound and vibration} \textbf{39}, 16–17 (2005).
\bibitem{Brandt2011} Brandt, A. \textit{Rotating Machinery Analysis. Signal analysis and experimental procedures} (John Wiley and Sons Ltd, Chichester, UK, 2011). 10.1002/9780470978160.ch12.
\bibitem{kateris2014} Kateris, D. \textit{et al.} A machine learning approach for the condition monitoring of rotating machinery. \textit{J Mech Sci Technol} \textbf{28}, 61–71; 10.1007/s12206-013-1102-y (2014).
\bibitem{wang2019c} Wang, Y. \textit{et al.} Order spectrogram visualization for rolling bearing fault detection under speed variation conditions. \textit{Mechanical Systems and Signal Processing} \textbf{122}, 580–596; 10.1016/j.ymssp.2018.12.037 (2019).
\bibitem{mcinerny2003} McInerny, S. A. \& Dai, Y. Basic vibration signal processing for bearing fault detection. \textit{IEEE Trans. Educ.} \textbf{46}, 149–156; 10.1109/TE.2002.808234 (2003). 
\bibitem{randall2011} Randall, R. B. \& Antoni, J. Rolling element bearing diagnostics—A tutorial. \textit{Mechanical Systems and Signal Processing} \textbf{25}, 485–520; 10.1016/j.ymssp.2010.07.017 (2011).
\bibitem{sun2018} Sun, J., Yan, C. \& Wen, J. Intelligent Bearing Fault Diagnosis Method Combining Compressed Data Acquisition and Deep Learning. \textit{IEEE Trans. Instrum. Meas.} \textbf{67}, 185–195; 10.1109/TIM.2017.2759418 (2018).
\bibitem{liu2016} Liu, H., Li, L. \& Ma, J. Rolling Bearing Fault Diagnosis Based on STFT-Deep Learning and Sound Signals. \textit{Shock and Vibration} \textbf{2016}, 1–12; 10.1155/2016/6127479 (2016).
\bibitem{liu2018} Liu, R., Yang, B., Zio, E. \& Chen, X. Artificial intelligence for fault diagnosis of rotating machinery: A review. \textit{Mechanical Systems and Signal Processing} \textbf{108}, 33–47; 10.1016/j.ymssp.2018.02.016 (2018).
\bibitem{zhao2019} Zhao, R. \textit{et al.} Deep learning and its applications to machine health monitoring. \textit{Mechanical Systems and Signal Processing} \textbf{115}, 213–237; 10.1016/j.ymssp.2018.05.050 (2019).
\bibitem{mey2020} Mey, O., Neudeck, W., Schneider, A. \& Enge-Rosenblatt, O. Machine Learning-Based Unbalance Detection of a Rotating Shaft Using Vibration Data. In \textit{2020 25th IEEE International Conference on Emerging Technologies and Factory Automation (ETFA)} (IEEETuesday, September 8, 2020 - Friday, September 11, 2020), pp. 1610–1617.
\bibitem{serin2020} Serin, G., Sener, B., Ozbayoglu, A. M. \& Unver, H. O. Review of tool condition monitoring in machining and opportunities for deep learning. \textit{Int J Adv Manuf Technol} \textbf{109}, 953–974; 10.1007/s00170-020-05449-w (2020).
\bibitem{zhang2020} Zhang, S., Zhang, S., Wang, B. \& Habetler, T. G. Deep Learning Algorithms for Bearing Fault Diagnostics—A Comprehensive Review. \textit{IEEE Access} \textbf{8}, 29857–29881; 10.1109/ACCESS.2020.2972859 (2020).
\bibitem{mey2021} Mey, O. \textit{et al.} Condition Monitoring of Drive Trains by Data Fusion of Acoustic Emission and Vibration Sensors. \textit{Processes} \textbf{9}, 1108; 10.3390/pr9071108 (2021).
\bibitem{adadi2018} Adadi, A. \& Berrada, M. Peeking Inside the Black-Box: A Survey on Explainable Artificial Intelligence (XAI). \textit{IEEE Access} \textbf{6}, 52138–52160; 10.1109/ACCESS.2018.2870052 (2018).
\bibitem{guidotti2019} Guidotti, R. \textit{et al.} A Survey of Methods for Explaining Black Box Models. \textit{ACM Comput. Surv.} \textbf{51}, 1–42; 10.1145/3236009 (2019).
\bibitem{jeyakumar2020} Jeyakumar, J. V., Noor, J., Cheng, Y.-H., Garcia, L. \& Srivastava, M. How Can I Explain This to You? An Empirical Study of Deep Neural Network Explanation Methods. In \textit{Advances in Neural Information Processing Systems}, edited by H. Larochelle, M. Ranzato, R. Hadsell, M. F. Balcan \& H. Lin (Curran Associates, Inc2020), Vol. 33, pp. 4211–4222.
\bibitem{zhang2018} Zhang, Q. \& Zhu, S. Visual interpretability for deep learning: a survey. \textit{Frontiers Inf Technol Electronic Eng} \textbf{19}, 27–39; 10.1631/FITEE.1700808 (2018).
\bibitem{selvaraju2017} Selvaraju, R. R. \textit{et al.} Grad-CAM: Visual Explanations from Deep Networks via Gradient-Based Localization. In \textit{2017 IEEE International Conference on Computer Vision. ICCV 2017 : proceedings : 22-29 October 2017, Venice, Italy} (IEEE, Piscataway, NJ, 2017), pp. 618–626.
\bibitem{bach2015} Bach, S. \textit{et al.} On Pixel-Wise Explanations for Non-Linear Classifier Decisions by Layer-Wise Relevance Propagation. \textit{PloS one} \textbf{10}, e0130140; 10.1371/journal.pone.0130140 (2015).
\bibitem{ribeiro2016} Ribeiro, M. T., Singh, S. \& Guestrin, C. "Why Should I Trust You?". \textit{In Proceedings of the 22nd ACM SIGKDD International Conference on Knowledge Discovery and Data Mining}, edited by B. Krishnapuram, \textit{et al.} (ACM, New York, NY, USA, 08132016), pp. 1135–1144.
\bibitem{shrikumar2017} Shrikumar, A., Greenside, P. \& Kundaje, A. Learning Important Features Through Propagating Activation Differences. In \textit{Proceedings of the 34th International Conference on Machine Learning}, edited by D. Precup \& Y. W. Teh (PMLR2017), Vol. 70, pp. 3145–3153.
\bibitem{sundara2017} Sundararajan, M., Taly, A. \& Yan, Q. Axiomatic Attribution for Deep Networks. In \textit{Proceedings of the 34th International Conference on Machine Learning}, edited by D. Precup \& Y. W. Teh (PMLR2017), Vol. 70, pp. 3319–3328.
\bibitem{lundberg2017} Lundberg, S. M. \& Lee, S.-I. A Unified Approach to Interpreting Model Predictions. In \textit{Advances in Neural Information Processing Systems}, edited by I. Guyon, \textit{et al.} (Curran Associates, Inc2017), Vol. 30.
\bibitem{kindermans2019} Kindermans, P.-J. \textit{et al.} The (Un)reliability of Saliency Methods. In \textit{Explainable AI: Interpreting, Explaining and Visualizing Deep Learning}, edited by W. Samek, G. Montavon, A. Vedaldi, L. K. Hansen \& K.-R. M\"uller (Springer International Publishing, Cham, 2019), pp. 267–280.
\bibitem{nath2021} Nath, A. G., Udmale, S. S. \& Singh, S. K. Role of artificial intelligence in rotor fault diagnosis: a comprehensive review. \textit{Artif Intell Rev} \textbf{54}, 2609–2668; 10.1007/s10462-020-09910-w (2021).
\bibitem{chen2020} Chen, H.-Y. \& Lee, C.-H. Vibration Signals Analysis by Explainable Artificial Intelligence (XAI) Approach: Application on Bearing Faults Diagnosis. \textit{IEEE Access} \textbf{8}, 134246–134256; 10.1109/ACCESS.2020.3006491 (2020).
\bibitem{kim2020} Kim, J. \& Kim, J.-M. Bearing Fault Diagnosis Using Grad-CAM and Acoustic Emission Signals. \textit{Applied Sciences} \textbf{10}, 2050; 10.3390/app10062050 (2020).
\bibitem{lin2021} Lin, C.-J. \& Jhang, J.-Y. Bearing Fault Diagnosis Using a Grad-CAM-Based Convolutional Neuro-Fuzzy Network. \textit{Mathematics} \textbf{9}, 1502; 10.3390/math9131502 (2021).
\bibitem{saeki2019} Saeki, M., Ogata, J., Murakawa, M. \& Ogawa, T. Visual explanation of neural network based rotation machinery anomaly detection system. In \textit{2019 IEEE International Conference on Prognostics and Health Management (ICPHM)} (2019), pp. 1–4.
\bibitem{kim2021} Kim, M. S., Yun, J. P. \& Park, P. An Explainable Convolutional Neural Network for Fault Diagnosis in Linear Motion Guide. \textit{IEEE Trans. Ind. Inf.} \textbf{17}, 4036–4045; 10.1109/TII.2020.3012989 (2021).
\bibitem{yoo2022} Yoo, Y. \& Jeong, S. Vibration analysis process based on spectrogram using gradient class activation map with selection process of CNN model and feature layer. \textit{Displays} \textbf{73}, 102233; 10.1016/j.displa.2022.102233 (2022).
\bibitem{liu2022} Liu, C., Meerten, Y., Declercq, K. \& Gryllias, K. Vibration-based gear continuous generating grinding fault classification and interpretation with deep convolutional neural network. \textit{Journal of Manufacturing Processes} \textbf{79}, 688–704; 10.1016/j.jmapro.2022.04.068 (2022).
\bibitem{kim2021b} Kim, M. S., Yun, J. P. \& Park, P. An Explainable Neural Network for Fault Diagnosis With a Frequency Activation Map. \textit{IEEE Access} \textbf{9}, 98962–98972; 10.1109/ACCESS.2021.3095565 (2021).
\bibitem{grezmak2019} Grezmak, J., Wang, P., Sun, C. \& Gao, R. X. Explainable Convolutional Neural Network for Gearbox Fault Diagnosis. \textit{Procedia CIRP} \textbf{80}, 476–481; 10.1016/j.procir.2018.12.008 (2019).
\bibitem{grezmak2020} Grezmak, J., Zhang, J., Wang, P. \& Gao, R. X. Multi-stream convolutional neural network-based fault diagnosis for variable frequency drives in sustainable manufacturing systems. \textit{Procedia Manufacturing} \textbf{43}, 511–518; 10.1016/j.promfg.2020.02.181 (2020).
\bibitem{hasan2021} Hasan, M. J., Sohaib, M. \& Kim, J.-M. An Explainable AI-Based Fault Diagnosis Model for Bearings. \textit{Sensors (Basel, Switzerland)} \textbf{21}; 10.3390/s21124070 (2021).
\bibitem{li2019} Li, X., Zhang, W. \& Ding, Q. Understanding and improving deep learning-based rolling bearing fault diagnosis with attention mechanism. \textit{Signal Processing} \textbf{161}, 136–154; 10.1016/j.sigpro.2019.03.019 (2019).
\bibitem{wang2020b} Wang, H., Liu, Z., Peng, D. \& Qin, Y. Understanding and Learning Discriminant Features based on Multiattention 1DCNN for Wheelset Bearing Fault Diagnosis. \textit{IEEE Trans. Ind. Inf.} \textbf{16}, 5735–5745; 10.1109/TII.2019.2955540 (2020).
\bibitem{fordatis2020} Mey, O., Neudeck, W., Schneider, A. \& Enge-Rosenblatt, O. Vibration Measurements on a Rotating Shaft at Different Unbalance Strengths. \textit{Fordatis}; 10.24406/fordatis/65.2 (2020).
\bibitem{our_github} Supplementary Information: Source Code Documentation of This Paper at Github. Available online: \nolinkurl{https://github.com/o-mey/xai-vibration-fault-detection}.
\bibitem{zhou2016} Zhou, B., Khosla, A., Lapedriza, A., Oliva, A. \& Torralba, A. Learning Deep Features for Discriminative Localization. In \textit{29th IEEE Conference on Computer Vision and Pattern Recognition. CVPR 2016 : proceedings : 26 June-1 July 2016, Las Vegas, Nevada} (IEEE, Piscataway, NJ, 2016), pp. 2921–2929.
\bibitem{chattopadhay2017} Chattopadhay, A., Sarkar, A., Howlader, P. \& Balasubramanian, V. N. Grad-CAM++: Generalized Gradient-Based Visual Explanations for Deep Convolutional Networks. In \textit{2018 IEEE Winter Conference on Applications of Computer Vision - WACV 2018. Proceedings : 12-15 March 2017, Lake Tahoe, Nevada} (IEEE, Piscataway, NJ, 2018), pp. 839–847.
\bibitem{wang2020} Wang, H. \textit{et al.} Score-CAM: Score-Weighted Visual Explanations for Convolutional Neural Networks. In \textit{2020 IEEE/CVF Conference on Computer Vision and Pattern Recognition Workshops (CVPRW)} (IEEESunday, June 14, 2020 - Friday, June 19, 2020), pp. 111–119.
\bibitem{alber2019} Alber, M. \textit{et al.} iNNvestigate Neural Networks! \textit{Journal of Machine Learning Research} \textbf{20}, 1–8 (2019).
\bibitem{lime4time} Emanuel Metzenthin. LIME For Time. Available online: \nolinkurl{https://github.com/emanuel-metzenthin/Lime-For-Time} (Accessed 2022-07-11).
\bibitem{viridis} S. Garnier, N. Ross, R. Rudis, A. P. Camargo, M. Sciaini, and C. Scherer, "viridis - Colorblind-Friendly Color Maps for R," R package version 0.6.2,  {https://sjmgarnier.github.io/viridis/}, doi = {10.5281/zenodo.4679424 }, 2021
\end{thebibliography}
\end{document}